\documentclass[aps, pre, amsmath, amssymb, showpacs,showkeys,floatfix,superscriptaddress,twocolumn]{revtex4}

\usepackage{natbib}
\usepackage{graphicx}
\usepackage{bm}
\usepackage{color}
\usepackage{epic, eepic}
\usepackage{subfigure}

\newcommand{\ddt}[1] {\partial_t {#1} }
\newcommand{\ddx}[2] {\partial_{#1} {#2}}
\newcommand{\ddxi}[1] {\partial_i {#1}}
\newcommand{\ddxj}[1] {\partial_j {#1}}
\newcommand{\ddxsq}[2] {\partial_{#1}^2 {#2} }
\newcommand{\ddxjsq}[1] {\partial_j^2 {#1} }

\newcommand{\av}[1] {\overline{#1}}
\newcommand{\fl}[1] {{#1}'}

\newcommand{\hav}[1] {\left< {#1} \right>_z}

\newcommand{\flt} {\mathcal{L}}
\newcommand{\proj} {P}
\newcommand{\projflt} {\mathcal{G}}
\newcommand{\fou}[1] {\widehat{#1}}
\newcommand{\sm}[1] {\widetilde{#1}}

\newcommand{\bP} {\mathcal{P}}
\newcommand{\bD} {\varepsilon}
\newcommand{\bT} {\mathcal{T}}

\renewcommand{\vec}[1] {\mathbf{#1}}
\newcommand{\ignore}[1] {}

\renewcommand{\Re} {\textrm{Re}}
\newcommand{\Ra} {\textrm{Ra}}
\renewcommand{\Pr} {\textrm{Pr}}
\newcommand{\Nu} {\textrm{Nu}}

\begin{document}

\title{Leray-alpha simulations of wall-bounded turbulent flows}

\author{Maarten \surname{van Reeuwijk}}
\affiliation{Department of Civil and Environmental Engineering, Imperial College London, Imperial College Road, London, SW7 2AZ, UK}
\email{m.vanreeuwijk@imperial.ac.uk}

\author{Harm J.J. \surname{Jonker}}
\affiliation{Department of Multi-Scale Physics and
             J.M. Burgers Center for Fluid Dynamics,
             Delft University of Technology,
             Lorentzweg 1, 2628 CJ Delft, The Netherlands}

\author{Kemo \surname{Hanjali\'{c}}}
\affiliation{Department of Multi-Scale Physics and
             J.M. Burgers Center for Fluid Dynamics,
             Delft University of Technology,
             Lorentzweg 1, 2628 CJ Delft, The Netherlands}
\affiliation{Department of Mechanics and Aeronautics, University of Rome, ``La Sapienza'', Rome, Italy}

\date{July 1, 2009}

\begin{abstract}
The Leray-$\alpha$ model reduces the range of active scales of the Navier-Stokes
equations by smoothing the advective transport.
Here we assess the potential of the Leray-$\alpha$ model in its standard formulation to simulate wall-bounded flows. Three flow cases are considered: plane channel flow at $\Re_\tau=590$, Rayleigh-B\'{e}nard convection at $\Ra=10^7$
and $\Pr=1$, and a side-heated vertical channel at $\Ra=5 \times 10^6$ and $\Pr=0.7$.  
The simulations are compared to results from a well resolved and coarse DNS.
It is found that for all three flow cases, the variance in the velocity
field increases as the filter width parameter $a$ is increased, where $a$ is
connected to the filter width as $\alpha_i = a \Delta x_i$, with $\Delta
x_i$ the local grid size.
Furthermore, the viscous and diffusive wall regions tend to thicken relative to the coarse DNS
results as a function of $a$.
In the cases where coarse DNS overpredicts wall gradients (as for
Rayleigh-B\'{e}nard convection and the side-heated vertical channel), the
thickening is beneficial.
However, for the plane channel flow, coarse DNS underpredicts the wall-shear
velocity, and increasing $a$ only degrades the results.
It is shown that buoyancy effects need to be included with care, because of the close relation of turbulent heat flux and the production of turbulent kinetic energy.
\end{abstract}

\maketitle

\section{Introduction}

The usual way to surmount the computational complexity of high Reynolds number turbulent flows and the exorbitant grid resolution demands of high Reynolds number turbulent flows is to average or filter the Navier-Stokes equations.
With such pre-processing of the equations, one ends up with unclosed stresses which need to be modeled.
In large-eddy simulations, the unclosed stresses are usually modeled with eddy-diffusivity type models.
An alternative way of modeling is to smooth the dynamics of the Navier-Stokes equations by a direct modification of the advective terms \citep{Holm1999, Foias2001, Cheskidov2005, Geurts2006a}.
This allows one to derive the corresponding subscale model without any
further assumptions, provided the employed filter $\flt$ is invertible.
The modification can be chosen such that important physical properties of the
Navier-Stokes equations are preserved, such as energy, helicity, the Kelvin
circulation theorem, or symmetries.

The Leray-$\alpha$ model \citep{Cheskidov2005, Geurts2006a,
vanReeuwijk2005a} is the simplest regularization model. For this model, the
advective operator $\ddxj{u_j u_i}$ is replaced by $\ddxj{\sm{u}_j u_i}$, where the
filtered velocity $\sm{u}_i = \flt(u_i)$ and $\flt$ is a filter operation.
Historically, this model was used to prove
existence and regularity of $u_i$, as well as convergence to the
Navier-Stokes solution as the filter width tends to zero \citep{Leray1934}.
Analytical estimates invoking Prandtl-Blasius arguments \citep{Cheskidov2005}
predict an excellent match for the laminar, transitional and turbulent
regime, although the physical relevance of the solutions in the turbulent
regime with respect to the boundary conditions and turbulence intensities may be questioned \citep{vanReeuwijk2007}.
The Leray-$\alpha$ model has been used for the simulation of the turbulent
mixing layer, and showed superior accuracy in comparison with dynamic eddy-viscosity models, in particular with respect to the quality of the small resolved scales \citep{Geurts2006a}.
The explicit filtering concept in LES can be traced back to \cite{Clark1979}, although that model was originally proposed by \cite{Leonard1974}.

The Lagrangian averaged Navier-Stokes-$\alpha$ (LANS-$\alpha$) model
\citep{Holm1998, Chen1999, Chen1999a, Chen1999b, Holm1999, Foias2001, Holm2002,
Mohseni2003, Marsden2003} in addition, posesses a filtered Kelvin circulation
theorem.
The LANS-$\alpha$ model has been successful in simulations of homogeneous
isotropic turbulence \citep{Chen1999a, Mohseni2003} and for the temporal mixing
layer \citep{Geurts2006a}.
A regularization approach which preserves the skew-symmetry of the advective
operator has been recently proposed by \cite{Verstappen2007}.
Here, the energy, enstrophy (in 2D) and helicity are conserved, and accurate
results are obtained with coarse simulations of plane channel flow.

Given the excellent results obtained with the Leray-alpha model the turbulent mixing layer \citep{Geurts2006a}, it is a natural step to assess whether the Leray-alpha model performs as well for wall-bounded flows. In this paper, we study three canonical flow cases which represent a wide range of problems occuring in practical situations. The first is a plane channel flow, which is one of the standard test cases for turbulence models. The main production mechanism of turbulent kinetic energy for this flow is by shear production. The second test case is Rayleigh-B\'{e}nard convection, for which the turbulence is created purely by buoyancy production. This is a challenging test case because of the sensitive dependence of the heat-transfer on the turbulent dynamics. The third test case is a side heated vertical channel, for which turbulent kinetic energy is produced both by shear and buoyancy. This test-case is challenging because it exhibits counter-gradient turbulence heat and momentum fluxes \citep{Hanjalic2002}. 


In this study, we perform a coarse DNS simulation for each flow case. The resolution for this simulation is chosen such that the turbulence statistics have degraded relative to the resolved DNS solution, but the numerical solution is not dominated by wiggles. This simulation is then taken as the reference. The expectation is that the Leray-alpha model will improve on these results by modifying the advective transport. The filter used to smooth the transport velocity is associated with the Helmholtz operator, as proposed in the original formulation of the Leray-$\alpha$ model \citep{Cheskidov2005}. In order to avoid excessive smoothing in the viscous and diffusive wall regions,
a non-uniform filter width is used in the wall-normal direction, which is related to the local gridsize. The non-uniform filtering gives rise to a nonsolenoidal filtered velocity field. As shown in \cite{vanReeuwijk2005a}, a non-solenoidal velocity field gives rise to spurious production of TKE very close to the wall, which degrades the performance of the Leray-$\alpha$ model. In order to circumvent this issue, a projection method is used to ensure that the filtered velocity be solenoidal.

The details of the Leray-$\alpha$ model are
outlined in section \ref{par:leray}.
A preliminary study \citep{vanReeuwijk2005a} showed that for Rayleigh-B\'enard convection, filtering with a
constant filter size leads to an artificial thickening of the hydrodynamic and thermal boundary layer
and significantly increased variances.
As properly resolving viscous and diffusive wall region is of primary importance for
wall-bounded flows, a space-varying filter size is applied in the wall-normal
direction which scales with the grid resolution.
This has implications for the velocity  field, and the details about enforcing a
divergence-free smoothed velocity field $\sm{u}_i$ when using a
space-varying filter \citep{vanReeuwijk2006} are discussed in section \ref{par:divergencefree}.
The simulations for the channel flow, Rayleigh-B\'{e}nard convection and
the side-heated convection are discussed in section \ref{par:results}.
The simulation of buoyancy-driven flows requires an extension of the
momentum equation with the buoyancy force and an additional temperature
equation.
It turns out for the adopted formulation, the direct coupling between the
turbulent heat-flux and the buoyancy production of turbulent kinetic energy is
lost, which is particularly problematic for Rayleigh-B\'{e}nard convection
(section \ref{par:discussion} and appendix \ref{app:rbexact}).
Several suggestions are presented to correct for this undesired side effect.
Concluding remarks are made in section \ref{par:conclusions}.

\section{\label{par:leray}Leray-$\alpha$ regularization}

The governing equations for the Leray-$\alpha$ model are 
\begin{gather}
\label{eq:leray}
\ddt{u_i} + \sm{u}_j  \ddxj{u_i} = \nu \ddxjsq{u_i} - \ddxi{p} + f_i, \\
\label{eq:div}
\ddxi{u_i} = 0, \\
\label{eq:alpha}
(1 - \ddxj{}\alpha_j^2\ddxj{})\sm{u}_i = u_i,
\end{gather}
with $u_i$ the velocity, $p$ the pressure, $\nu$ the kinematic viscosity and
$f_i$ a body force.
As can be seen in (\ref{eq:leray}), the regularization 
modeling approach results in a mixed formulation of filtered and unfiltered
quantities.
Even though $u_i$ is unfiltered it should not be considered a velocity coming from direct
numerical simulation (DNS) because of the modified nonlinearity in
\eqref{eq:leray}.
Therefore, both $u_i$ and $\sm{u}_i$ are regularized variables.
The variable $\alpha_i$ corresponds to the filter width which can have
different values per direction and varies in the wall-normal direction.
The filter $\flt$ is given by the inverse of the elliptic equation
(\ref{eq:alpha}) and boundary conditions; we will denote the 
filtering operation as
\begin{equation}
\label{eq:flt}
\sm{u}_i = \flt(u_i) = (1 - \ddxj{}\alpha_j^2\ddxj{})^{-1}\ u_i.
\end{equation}
This is the standard filter used for analytical studies of the Leray-$\alpha$
and LANS-$\alpha$ model \citep{Foias2001, Cheskidov2005}.
Naturally, other filters could be used as well, as the effect of the filtering
would (hopefully) be restricted to the small scales only.
One advantage of using a formulation like (\ref{eq:alpha}) is 
in the natural treatment of boundaries, as explicit filters require special treatment near the boundary.
However, solving 
(\ref{eq:alpha}) is quite costly, and care should be taken to ensure the incompressibility of the filtered velocity $\sm{u}$ \citep{vanReeuwijk2006}.

The Leray-$\alpha$ model does not add additional dissipation to the equations.
Indeed, the equation can be written in an LES template as 
\begin{equation}
  \label{eq:NStemplate}
  \ddt u_i + \ddxj{u_j u_i} + \ddxi{p} - \nu \ddxjsq{u_i} = \ddxj{m_{ij}},
\end{equation}
where $m_{ij}$ is the subfilter model, which is given by $m_{ij} = u''_j u_i$.
Here, $u''_j= u_j - \sm{u}_j$ is the high wavenumber part of $u_i$.
Multiplying $m_{ij}$ with $u_i$ results in 
$u_i \ddxj{m_{ij}} = \ddxj{(u''_j \frac{1}{2}u_i^2)}$, which shows that energy is only redistributed and not dissipated.

One can also derive the implied subfilter scale model of the Leray-$\alpha$
model in terms of the filtered velocity $\sm{u}$.
Filtering (\ref{eq:leray}) with $\flt$ leads to an explicitly filtered
LES template, given by
\begin{equation}
  \label{eq:LEStemplate}
  \ddt{\sm{u}_i} + \ddxj{\sm{u}_j \sm{u}_i} + \ddxi{\sm{p}} 
    - \nu \ddxjsq{\sm{u_i}} = \ddxj{\sm{m}_{ij}},
\end{equation}
but with a non-standard asymmetric subfilter term
\begin{equation}
  \sm{m}_{ij} = \sm{u}_j \sm{u}_i - \sm{\sm{u}_j u_i}
         = \sm{u}_j \sm{u}_i - \sm{u_j u_i} + \sm{u''_j u_i}
\end{equation}

\begin{figure}
\centering
\includegraphics[width=60mm]{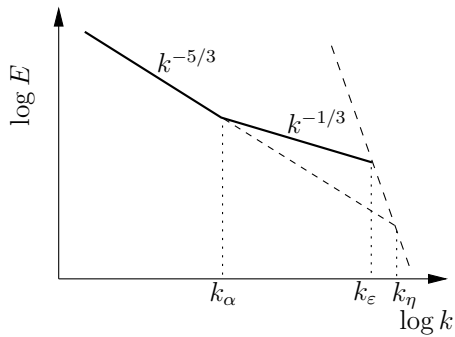}
\caption{\label{fig:spectrum} Sketch of a typical spectrum for the
unfiltered velocity $u_i$ of the Leray-$\alpha$ model.}
\end{figure}

The Leray-$\alpha$ model reduces the number of active scales by slowing down the energy cascade at small scales. This is shown schematically in Fig.\ \ref{fig:spectrum} for the energy spectrum $E$ of unfiltered velocity. The wavenumber associated with the filter size is $k_\alpha$. For $k<k_\alpha$, the energy cascade is unaffected which results in the classical $E \propto k^{-5/3}$ spectrum. However, for $k>k_\alpha$, the filter affects the cascade time-scale and the spectrum is modified to $E \propto k^{-1/3}$. Because of the higher variances at small scales, the dissipation will take place on larger scales than the Kolmogorov scale with associated wave number $k_\eta$. 
Note that the energy density spectrum of the filtered velocity
$\sm{E}=\fou{\flt}^2(k) E(k)$ falls off steeply as $\sm{E} \propto k^{-13/3}$ beyond $k_\alpha$ \citep[see also][]{Cheskidov2005}. 

\section{\label{par:divergencefree}Enforcing a nondivergent $\sm{u}_i$ field}

Enforcing the incompressibility condition $\ddxj{u}_j=0$ does not automatically
ensure that $\ddxj{\sm{u}}_j=0$ for a wall-bounded flow. For uniform $\alpha$, the solution  $f(\vec{x})=\ddxj{\sm{u}_i}$ of the homogeneous differential equation $f-\ddxj{\alpha_j^2\ddxj{}}f = 0$ is nontrivial and does not vanish upon applying the boundary conditions \citep{vanReeuwijk2006}. 
The main problem with $\ddxj{\sm{u}_j} \neq 0$ is that the purely redistributive
character of advection is lost, which can be demonstrated by considering the
effect on the balance of the turbulent kinetic energy (TKE).
The turbulent transport of TKE by fluctuations can be calculated by taking 
the advection term $\ddxj{ ( \fl{\sm{u}_j} \fl{u}_i )}$ from the equation of
velocity fluctuations , multiplying it by $\fl{u_i}$ and averaging:
\begin{equation}
\av{\fl{u}_i \ddxj{ ( \fl{\sm{u}_j} \fl{u}_i ) }} =
  \ddxj{\av{\fl{\sm{u}_j} \fl{e}}} +
  \av{\fl{e} \ddxj{\fl{\sm{u}_j}}},
\end{equation}
with $\fl{e} = \frac{1}{2} \fl{u}_i \fl{u}_i$.
The first term on the right-hand side is in divergence-form and hence purely
redistributive.
The second term is a production/destruction term, which normally
vanishes because the fluctuating field is divergence free.
However, when $\ddxj{\sm{u}_j} \neq 0$ this term can become nonzero, thereby
allowing for spurious TKE production and destruction.
As was shown for Rayleigh-B\'{e}nard convection at
$\Ra=10^5$ and $\Pr=1$, the term $\av{\fl{e} \ddxj{\fl{\sm{u}_j}}}$ is
responsible for significant energy production in the
hydrodynamic boundary layer (up to 25\% of the volume-averaged production)\citep{vanReeuwijk2006}.

A nonuniform filter width $\alpha$, creates another opportunity for $\ddxj{\sm{u}}_j\neq 0$.
Applying the filter $\flt$ to the divergence of the unfiltered velocity gives
that $\flt(\ddxj{u_j}) = \ddxj{\flt(u_j)} + [\flt, \ddxj{}] u_j$, where the
brackets represent the commutator defined as $[A,B] \equiv AB - BA$.
Therefore, the divergence of the filtered velocity is given by
\begin{equation}
   \label{eq:divergence}
   \ddxj{\sm{u_j}} = [\ddxj{}, \flt] u_j + \flt(0)
\end{equation}
This shows both ways in which $\sm{u}$ can cease to be divergence-free.
First the commutator may not vanish because of a nonuniform filter width
$\alpha(\vec{x})$.
Second, the term $\flt(0)$ is nonzero when the solution  $f(\vec{x})=\ddxj{\sm{u}_i}$ of the homogeneous differential equation $f-\ddxj{\alpha_j^2\ddxj{}}f = 0$ does not yield $f=0$ after applying the boundary conditions, as discussed above. 
In \cite{vanReeuwijk2006}, several methods were proposed to maintain a nondivergent $\sm{u}_i$ field: enforcing $\ddxj{\sm{u}_j}=0$ directly (and thus dropping $\ddxj{u_j}=0$), letting $\alpha$ vanish at the wall, using free-slip boundary conditions for $\sm{u}_i$ and introducing a projection method.
Of these four options, only the projection method can be successfully
applied for nonuniform filters $\alpha(\vec{x})$, so this is a convenient
choice.
The projection method \citep[e.g.][]{Ferziger2002} adds a gradient $\ddxi{\phi}$ to (\ref{eq:flt})
as $\sm{u_i}+\ddxi{\phi}=\flt(u_i)$ and solves a Poisson equation
$\ddxjsq{\phi} = \ddxi{\flt(u_i)}$ to make $\sm{u_i}$ incompressible.
The boundary conditions imposed on $\sm{u}_i$ and $\phi$ are equivalent to the conditions imposed on $u_i$ and $p$;  $\sm{u}_i=0$ and $\partial_n \phi=0$, respectively.

With the projection included, the filtered velocity $\sm{u}_i$ is related to the unfiltered velocity 
$u_i$ by a filter $\projflt$ as
\begin{equation}
  \label{eq:projflt}
  \sm{u}_i = \projflt(\vec{u}) = \proj_{ij} \circ \flt(u_j),
\end{equation}
where $\proj_{ij}$ is the projection operator.
Note that $\projflt$ takes the complete velocity vector as its argument, as
opposed to $\flt$ which only requires one velocity component at a time.

The grid resolution is normally a good indication for the location of the
most demanding flow features.
Therefore, we couple the filter size $\alpha_i$ directly to the grid
resolution by a parameter $a$ as 
\begin{equation}
  \label{eq:adef}
  \alpha_i = a \Delta x_i.
\end{equation}
Note that the filter size is anisotropic.
In the wall-normal direction, the grid is clustered towards to wall so that the
filter is small near the wall and larger in the midplane. The smoothing parameter $a$ is varied between $0 < a < 1$, for which the typical filter width spans up to four grid cells. This can be shown by calculating the filter width $\Delta$ from \cite{Geurts2006a}:
\begin{equation}
\label{eq:typicalfilterwidth}
\frac{1}{\Delta} = \int_{-\infty}^{\infty} G_\alpha^2(x, x') d x',
\end{equation}

where $G_\alpha$ is a normalized filter kernel. The free-space Green's function associated with the Helmholz equation is
\begin{equation}
  G_\alpha(x, x') = \frac{1}{2 \alpha} \exp \left( - \frac{| x - x' |}{\alpha} \right)
\end{equation}
Substitution of the above expression and \eqref{eq:adef} into \eqref{eq:typicalfilterwidth} leads to $\Delta = 4 a \Delta x_i$.

The projection method has been implemented in our code for direct
simulation as a two-step process.
In the first step, the elliptic equation (\ref{eq:alpha}) is solved by a direct method
which takes advantage of the homogeneous directions.
In the second step, the projection method is used to make this field divergence
free.
The code is based on staggered finite differences and
uses second order central differences in space and a second order
Adams-Bashforth scheme in time \citep{vanReeuwijk2007}.
The code is fully parallellized and supports grid clustering in wall-normal
direction.
Special care has been taken to preserve the purely redistributive character
of the advection scheme on the nonuniform grid by using a symmetry-preserving
discretization \citep{Verstappen2003}.

\section{\label{par:results}Results}
\subsection{\label{par:channel}Plane channel flow}


\begin{figure*}
\centering
\includegraphics[width=150mm]{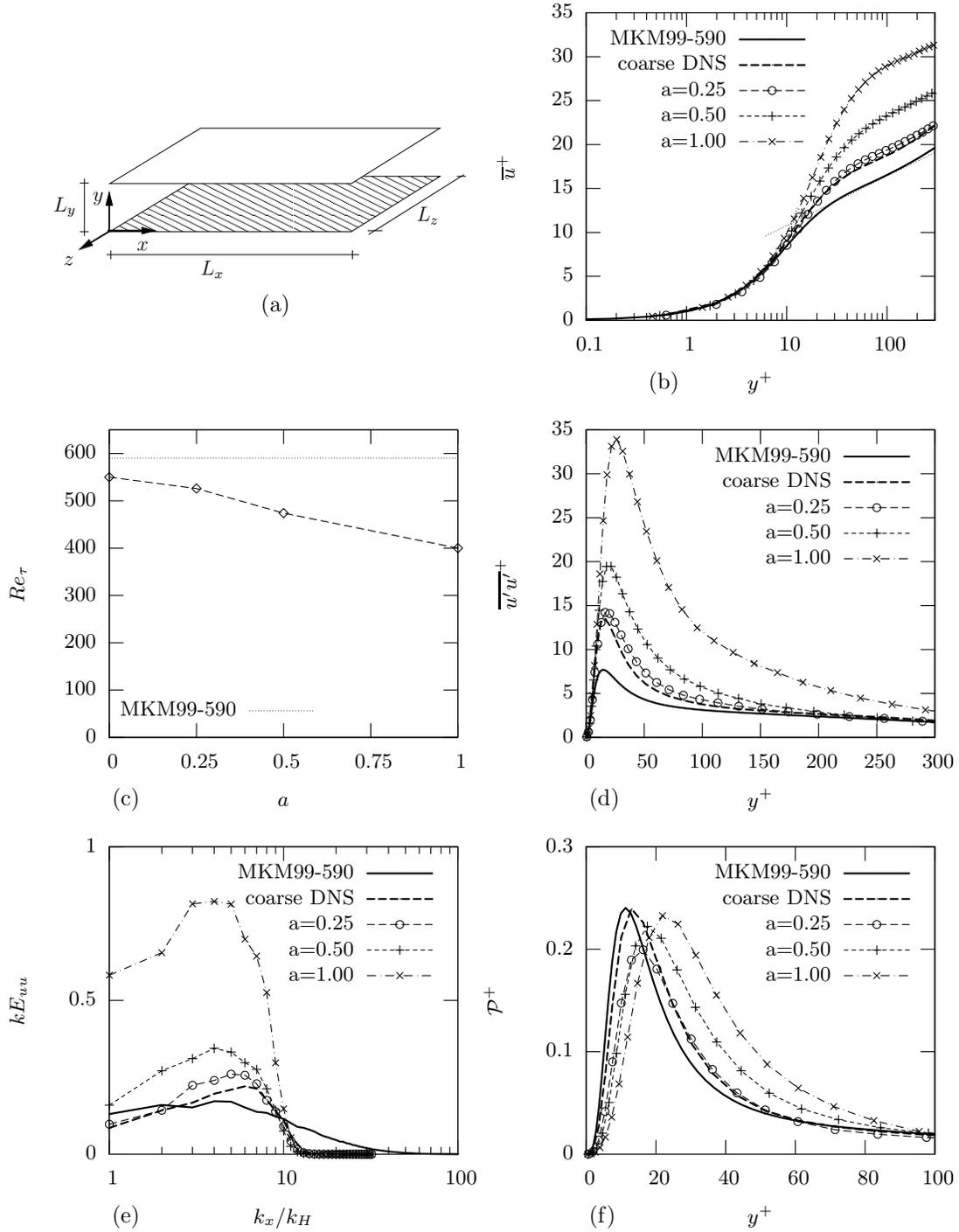}
\caption{\label{fig:channel}The effect of varying the filter size $a$ for plane channel flow. (a) definition sketch; (b) average velocity; (c) shear-Reynolds number $\Re_\tau$; (d) streamwise variance $\av{\fl{u}\fl{u}}$; (e) streamwise spectrum $E_{uu}$ in midplane; (f) production of turbulent kinetic energy.}
\end{figure*}

The first wall-bounded flow case under consideration is plane channel flow.
Here we compare results for the Leray-$\alpha$ model with direct simulations \citep{Moser1999} at $\Re_\tau=590$.
The coordinate system is defined in Fig.\ \ref{fig:channel}a, 
with the $x-$, $y-$ and $z$-coordinate aligned with the streamwise, wallnormal and
spanwise direction, respectively.
The medium is air with a kinematic viscosity $\nu=10^{-5} \text{m}^2\text{/s}$
and the domain size $L_x \times L_y \times L_z$ is 
$2 \pi \delta \times 2 \delta \times \pi \delta$ where $\delta= 1$ m is the
channel halfwidth.
No-slip boundary conditions are applied to the top and bottom plates,
and periodic boundary conditions are used on the sidewalls.
The average velocity is fixed at $U=0.115$ m/s for all simulations.
At full DNS resolution ($384\times 384 \times 256$) cells, this flowrate
results in a shear Reynolds number $\Re_\tau=610$.
For the Leray-$\alpha$ and coarse DNS simulations, the grid has been chosen such
that the grid is too coarse for a good solution without a subgrid model, but
fine enough so that the solution is not dominated by wiggles. 
For the plane channel flow this resulted in a grid of $64 \times 64 \times 32$ cells with strong
grid stretching (up to 16 percent near the walls).
All results were averaged for about 60 typical turnovers (based on a
typical turnover time $t^*=U/(2 \delta$).

\begin{figure}
\centering
\includegraphics{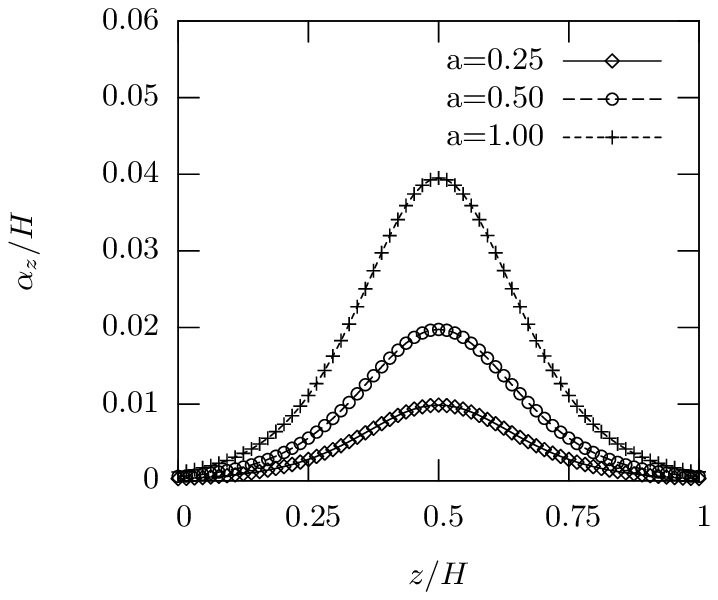}
\caption{\label{fig:chan_alpha} The distribution of $\alpha$ as a function of
$y$.}
\end{figure}

Three different values of the filter parameter $a$ are used
here: $a=0.25$, $0.50$ and $1.00$.
As discussed before, the filter is proportional to the local grid resolution as
$\alpha_i = a \Delta x_i$, and due to the grid clustering in the wall-normal
direction, the filter is much smaller near the wall than in the midplane (Fig.\
\ref{fig:chan_alpha}).
In this way, the artificial thickening of the viscous wall region \citep[viscous sublayer and buffer layer][]{Pope2000}. due to the
filtering which hampered preliminary simulations \citep{vanReeuwijk2005a} is
minimized.

Shown in Fig. \ref{fig:channel}b, are the average velocity profiles in 
plus-units, defined as $\av{u}^+=\av{u}/u_\tau$ and $y^+=y u_\tau / \nu$.
Here $u_\tau$ is the friction velocity defined as $u_\tau =
\sqrt{\tau_w/\rho}$, where $\tau_w=\mu \ddx{y}{\av{u}} |_w$ is the shear stress
at the wall.
As can be expected, the coarse DNS velocity profile lies above the reference
profile, which reflects an underestimated wall-shear stress.
The results for the Leray-$\alpha$ model are denoted by the circles, squares
and crosses for $a=0.25$, $0.50$ and $1.00$ respectively.
The simulation with $a=0.25$ has little influence on the average velocity,
but as $a$ increases the profiles deviate more and more from their desired
value.
The source of these deviations is a decreasing wall-shear stress.
Shown in Fig.\ \ref{fig:channel}c is the shear-Reynolds
number $\Re_\tau=u_\tau \delta / \nu$ as a function of $a$.
The dashed line corresponds to the expected value ($\Re_\tau=590$).
The coarse DNS results ($a=0$) show that the insufficient resolution
results in a 10\% underestimation of $\Re_\tau$.
The subfilter scale model would ideally compensate this effect.
However, as $a$ is increased, $\Re_\tau$ decreases up to an underestimation of
30\% at $a=1.00$.

The variance profile of the streamwise velocity component,
$\av{\fl{u}\fl{u}}$, is shown in
Fig.\ \ref{fig:channel}d, nondimensionalized by $u_\tau^2$.
The coarse DNS overpredicts the peak of $\av{\fl{u}\fl{u}}^+$ by a factor two, 
and increasing $a$ only makes
the situation worse, with a factor five overprediction at $a=1.00$. It should be noted that part of this increase is due to the normalisation: the decrease in the friction velocity causes $\av{\fl{u}\fl{u}}^+$ to become larger. However, even without scaling, the variance increases a factor 2.5 when comparing the resolved DNS results to the Leray-alpha simulation at $a=1$.

The peak in $\av{\fl{u}\fl{u}}^+$ can be seen to shift outwards, indicating a thickening of
the viscous wall region. The probable mechanism for a thickening of the viscous wall region resides in the modification of the momentum-flux $\av{\tilde{v}'u'}$ in the buffer layer. The filtering of $v$ will damp near-wall fluctuations, thereby reducing the turbulent momentum flux. As a result, the turbulence transport takes over further away from the wall than without filtering.

To get an idea of the typical lengthscales that cause the overprediction in
the variance, the spatial spectrum of $\av{\fl{u}\fl{u}}$ is presented.
The spectrum is obtained by one-dimensional Fast Fourier Transforms (FFT) in
the streamwise (spanwise) direction, and averaging over the other homogeneous
direction.
In addition, the spectrum is averaged over about 60 typical timescales to
eliminate slow transients.
In Fig.\ \ref{fig:channel}e, the spectrum of the 
streamwise velocity $u$ is shown in the midplane of the channel.
The coarse DNS
slightly underpredicts the variance on the large scales and overpredicts the
variance at the intermediate wavenumbers.
When $a$ is increased, first the variance increases at intermediate
wavenumbers.
For larger $a$, a significant increase in the variance occurs for the
large wavenumbers as well.
It should be noted that in the midplane, the difference between the DNS and
the Leray-$\alpha$ model is relatively small;
in the viscous wall region, the difference would be even greater. 

More insight into the enhanced variances may be obtained by studying the budget for turbulent kinetic energy (TKE).
The equation of TKE can be obtained by multiplying
the fluctuating part of (\ref{eq:leray}) by $u'_i$ and 
averaging over the homogeneous directions and over time.
The average $\av{\cdot}$ is denoted by an overline.
This results in
\begin{equation}
\label{eq:chan_tke}
0 = 
\underbrace{- \av{\fl{\sm{v}}\fl{u}} \ddx{y}{\av{u}}}_{\bP} - 
\underbrace{\nu 
 \av{(\ddxj{\fl{u}_i})(\ddxj{\fl{u}_i}})}_{\bD}
- \underbrace{\ddx{y}(\av{\fl{\sm{v}} e'} - \nu \ddx{y}{e} +
\av{\fl{v} \fl{p}})}_{\bT},
\end{equation}
where $e=\frac{1}{2} \av{\fl{u}_i\fl{u}_i}$ and $e' = \frac{1}{2}
\fl{u}_i\fl{u}_i$; $\bP$, $\bD$ and $\bT$ represent production, dissipation and
transport of turbulent kinetic energy respectively.
The smoothed transport velocity directly modifies the  shear production term $\bP$ and the transport of the velocity fluctuations $\av{\fl{\sm{v}} e'}$.

The fact that the variance increases while the production of TKE remains roughly the same is one of the intriguing features of the Leray-alpha model. Because of the attenuation of small-scale dynamics (Fig. \ref{fig:spectrum}), it is to be expected that the total variance will increase due to a slowing down of the cascade at high wavenumbers. A recent study \citep{Graham2007} suggests that the variance accumulation at subfilter scales is even higher; for the LANS-alpha model, the small scales seem to behave as "`rigid rotators"' which are advected passively by the larger scales.

The production of TKE is shown in Fig.
\ref{fig:channel}f, nondimensionalized by $u_\tau^4/\nu$.
The thickening of the viscous wall region is clear in the profile of $\bP^+$, where the peak (which occurs where the viscous stress is equal to the Reynolds stress) shifts from 
about $y^+=15$ for the coarse DNS to around $y^+=30$ for $a=1.00$.
Note also that the width of the production peak increases: only at $y^+=80$ is
$\bP$ at the value of the reference DNS for $a=1.00$.

\subsection{\label{par:rb}Rayleigh-B\'{e}nard convection}

\begin{figure*}
\centering
\includegraphics[width=150mm]{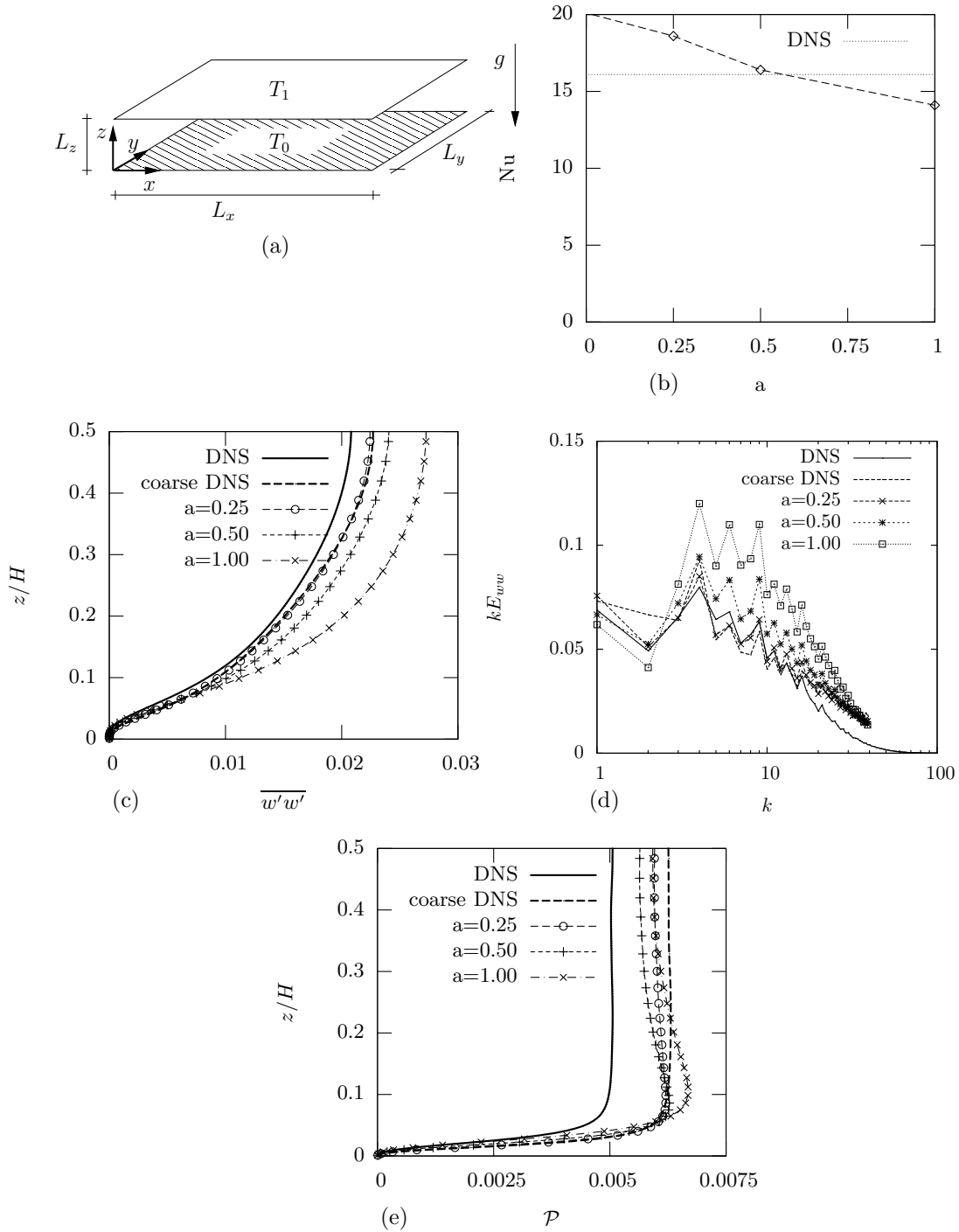}
\caption{\label{fig:rb}The effect of varying the filter size $a$ for Rayleigh-B\'{e}nard convection. (a) definition sketch; (b) Nusselt number $\Re_\tau$; (c) wall-normal variance $\av{\fl{w}\fl{w}}$; (d) wallnormal spectrum $E_{ww}$ in midplane; (f) production of turbulent kinetic energy.}
\end{figure*}

Rayleigh-B\'{e}nard convection (RB) is generated when a fluid in between two
flat plates is heated from the bottom and cooled from the top 
(Fig.\ \ref{fig:rb}a).
The system can be characterized by the Prandtl number $\Pr = \nu
\kappa^{-1}$ and the Rayleigh number $\Ra= \beta g \Delta \Theta H^3 (\nu
\kappa)^{-1}$.
The system reacts by a convective motion which is characterized by the
Reynolds number $\Re = U H \nu^{-1}$ and by an enhanced heat transfer through the
Nusselt number $\Nu = \phi H (\kappa \Delta \Theta)^{-1}$.
Here $U$ is a characteristic velocity and $\phi$ the realised heat-flux at the wall.
Both $\Re$ and $\Nu$ are non-trivial functions of $\Ra$ and $\Pr$ and are still the subject of ongoing research \citep[e.g.][]{Ahlers2009}.
The coordinate system is defined with the $z$-direction pointing upwards and the
gravity vector is in the negative $z$-direction.

The medium has most material properties of water except for the Prandtl number,
which is taken as  $\Pr=1$ instead of 7 to relax the computational
demands \citep{vanReeuwijk2008a}; the viscosity $\nu=1.07 \times 10^{-6}\
\text{m}^2/ \text{s}$, expansion coefficient $\beta=1.74 \times
10^{-4}\ K^{-1}$, $\Delta \Theta = 2\ K$, with $T_0=\Delta \Theta / 2$ and
$T_1=-\Delta \Theta / 2$.
The domain size $L_x \times L_y \times L_z$ is $\Gamma H \times \Gamma H
\times H$, with $H = 0.15$ m and the aspect ratio $\Gamma = 4$.
Fixed temperature and no-slip velocity boundary conditions are enforced at the
top and bottom plates. Periodic boundary conditions are used for the sidewalls.

The simulation of a buoyancy driven flow requires the extension of the Leray-alpha model with a transport equation for temperature.
This raises the question whether the advection of scalars should be with the
filtered or the unfiltered velocity.
Here the same modified advection operator will be used for all
transported quantities, i.e.\ advective transport with $\sm{u}_i$.
Invoking the Boussinesq approximation, the effect of buoyancy can be included 
by introducing a body force $\beta g_i \Theta$ in the momentum equations only.
After the extension with a transport equation of temperature and with the
additional body force, the Leray-$\alpha$ model becomes
\begin{gather}
\label{eq:leray2}
\ddt{u_i} + \sm{u}_j  \ddxj{u_i} = \nu \ddxjsq{u_i} - \ddxi{p}
- \beta g_i \Theta, \\
\label{eq:theta2}
\ddt{\Theta} + \sm{u}_j  \ddxj{\Theta} = \nu \ddxjsq{\Theta} 
\end{gather}

In the current coordinate system, the gravity is in the negative $z$-direction
as $g_i = - g \delta_{i3}$.
Due to the homogeneity and the absence of a forcing in the $x$ and $y$
direction, $\av{u}_i=0$, and the system is statistically one-dimensional.
\ignore{
Although this statement is mathematically correct, it does not give full justice
to the complexity of the Rayleigh-B\'{e}nard problem.
In reality, the system sets up a large scale circulation or 'wind'
\citep{Krishnamurti1981, Grossmann2000, Kadanoff2001} which is of great
importance to the heat transfer process and should be included in the analysis
\citep{vanReeuwijk2005, vanReeuwijk2008a}.
However, given the additional complexity of extracting the wind effects, we opt
for using the $\av{u}_i=0$ perspective at the cost of a deeper insight into the
changes of the flow physics.}
One of the characteristic features of RB is that the total vertical heat-flux
through the fluid is constant.
Averaging over the homogeneous directions, integrating (\ref{eq:theta2}) with
respect to $z$ and substituting the Dirichlet boundary conditions for
temperature results in
\begin{equation}
\label{eq:rb_heatflux}
\av{\fl{\sm{w}}\fl{\Theta}} - \kappa \ddx{z}{\av{\Theta}} = 
            \frac{\kappa \Delta \Theta}{H} Nu.
\end{equation}
Note that the turbulent heat-flux is in terms of
the filtered velocity $\sm{u}_i$.
However, the TKE production by buoyancy is given by $\av{\fl{w}\fl{\Theta}}$
(\ref{eq:rb_tke}) and this mismatch will prove to be of
importance, as discussed below.

Although many results can be found in the literature on direct simulation of
Rayleigh-B\'{e}nard convection in wide aspect ratios 
\citep{Domaradzki1994, Kerr1996}, no database is available as for plane
channel flow and side-heated convection simulations.
Therefore, the results will be compared to our own DNS results \citep{vanReeuwijk2005, vanReeuwijk2008a} at $\Ra=10^7$ and
$\Pr=1$.
The grid for the DNS is $256 \times 256 \times 256$ cells which is of
sufficient resolution not to require grid clustering near the wall.
The grid resolution for the Leray-$\alpha$ and coarse DNS simulations is $80
\times 80 \times 64$, which is again chosen such that it is too coarse for
DNS but is not yet dominated by numerical contamination.
Here, grid clustering is applied such that of the 64 wallnormal points, 8 are
present in each thermal boundary layer \footnote{Because most transport occurs perpendicular to the wall, the hydrodynamic and thermal boundary layers are commonly defined as the region from the wall to where the kinetic energy and temperature variance peaks, respectively.}

One of the most important integral flow properties in RB is the Nusselt number
$\Nu$.
At the current $\Ra$ and $\Pr$, the DNS gives that $\Nu = 16.1$.
The coarse DNS significantly overpredicts this value with $\Nu=20$ 
(Fig.\ \ref{fig:rb}b), due to the insufficient resolution.
The influence of the Leray-$\alpha$ model is to decrease $\Nu$, and at $a=0.50$
the Nusselt number $\Nu$ is approximately at its expected value.

In Fig.\ \ref{fig:rb}c, the wall-normal  velocity variance
$\av{\fl{w}\fl{w}}$ is shown, normalized by $U^2$
where $U=\sqrt{\beta g \Delta \Theta H}$ is the free-fall velocity.
As for the plane channel flow, the variance can be seen to increase as $a$ becomes larger.
Spatial spectra are collected by performing a 2D FFT and integrating
over circles $k_x^2+k_y^2 = k^2$.
Shown in Fig. \ref{fig:rb}d is the spectrum of the vertical velocity in
the midplane.
The increase of variance for the coarse DNS seems to be mainly concentrated at
the large scales (low wavenumbers).
When $a$ is increased, it can be seen that the variance at the intermediate
wavenumbers increases.
At the low wavenumbers, the trend is in the right direction.

As before, we study the equation of TKE, which is for Rayleigh-B\'{e}nard 
convection given by
\begin{equation}
\label{eq:rb_tke}
0 = 
\underbrace{\beta  g \av{\fl{w} \fl{\Theta}}}_{\bP} - 
\underbrace{\nu
 \av{(\ddxj{\fl{u}_i})(\ddxj{\fl{u}_i}})}_{\bD}
- \underbrace{\ddx{z}(\av{\fl{\sm{w}} e'} - \nu \ddx{z}{e} +
\av{\fl{w} \fl{p}})}_{\bT}. 
\end{equation}
For this flow case, there is no production on average of turbulent kinetic
energy by shear and the only effect of the filtering is in a modified transport
of the velocity fluctuations $\av{\fl{\sm{w}} e'}$.

The production term $\bP$ is shown in 
Fig.\ \ref{fig:rb}e, nondimensionalized by 
$(\beta g \Delta \Theta H)^{3/2}/H$.
The coarse DNS overestimates $\bP$, which is consistent with
the overestimation of $Nu$ as these are directly coupled for RB by the exact
relation \citep[see e.g.][]{Siggia1994, Grossmann2000} $\hav{\bP} = \hav{\bD} = \frac{\nu^3}{H^4} Ra (Nu -1) Pr^{-2}$.
The total production decreases for $a=0.25$ and $a=0.50$, but for $a=1.00$,
$\bP$ increases again.
This is not consistent with the monotonically decreasing trend for $Nu$ (Fig.
\ref{fig:rb}b).
Furthermore, the shape of the production profile changes: the constant
production in the bulk is replaced by a production profile which peaks
near the hydrodynamic boundary layers.
Both the trendbreak and the change of shape are the results of a disparity
between the buoyancy production term $\bP=\beta g \av{\fl{w} \fl{\Theta}}$ and
the turbulent heat-flux $\av{\fl{\sm{w}} \fl{\Theta}}$.
This is a fundamental issue which will be discussed in detail in section
\ref{par:discussion}.

\subsection{\label{par:lhc}Side-heated vertical channel}

\begin{figure*}
\centering
\includegraphics[width=150mm]{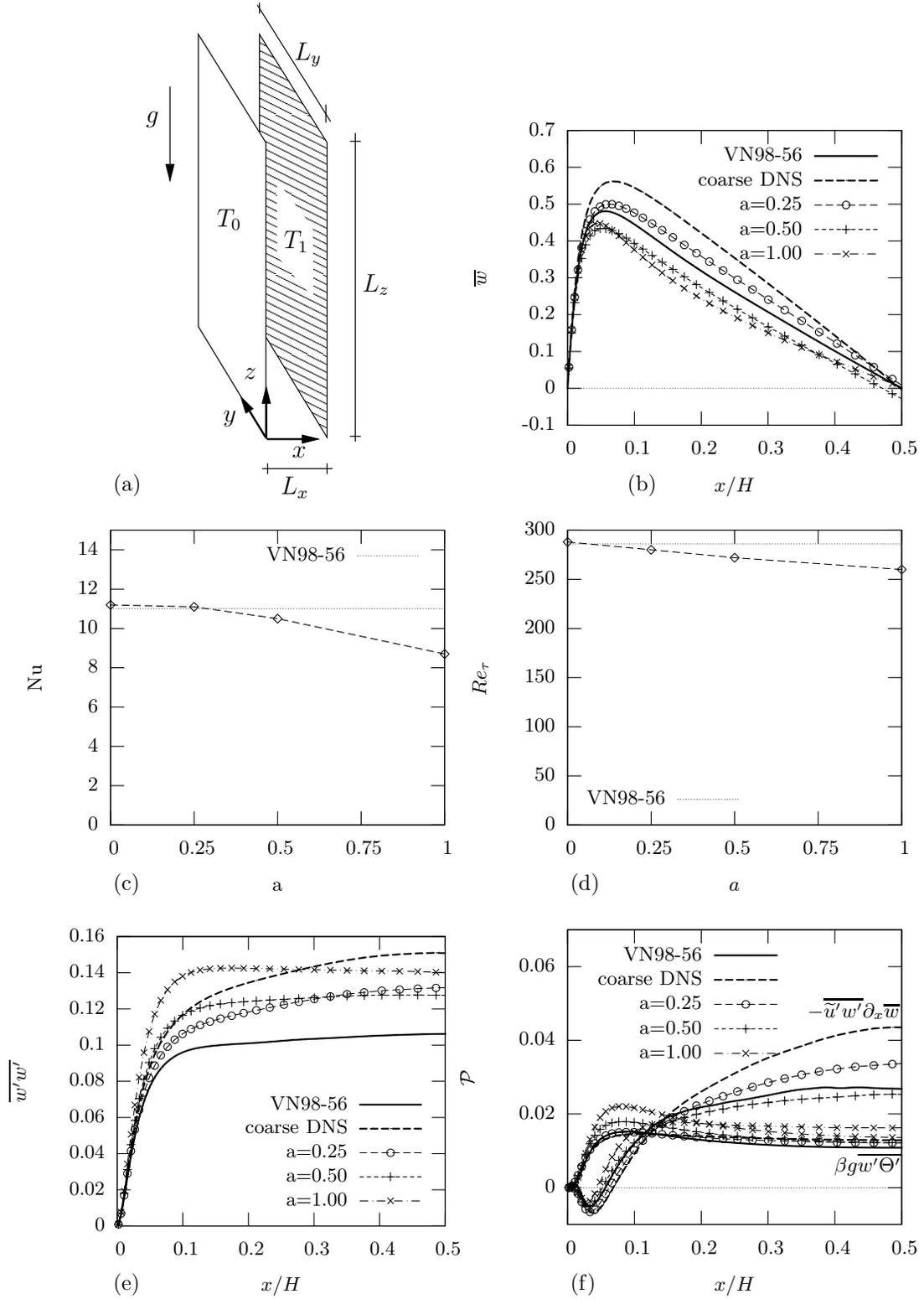}
\caption{\label{fig:lhc}The effect of varying the filter size $a$ for side-heated convection. (a) definition sketch; (b) average velocity; (c) Nusselt number $\Nu$; (d) shear-Reynolds number $\Re_\tau$; (e) streamwise variance $\av{\fl{w}\fl{w}}$; (f) production of turbulent kinetic energy.}
\end{figure*}

The side-heated vertical channel is a case where both buoyancy and shear
production are important.
The flow has several unusual features, such as negative shear-production
in the boundary layers \footnote{The hydrodynamic boundary layer can be determined using the maximum of $\av{w}$, and the thermal boundary layer by the maximum of the temperature variance.} and counter-gradient heat fluxes.
The side-heated vertical channel has been studied intensively both
experimentally \citep{Betts2000} and by direct
simulation 
\citep{Boudjemadi1997, Versteegh1998a, Versteegh1998, Versteegh1999}.
The flow geometry for the side-heated vertical channel is sketched in Fig.\
\ref{fig:lhc}a.
For this case, the $x$-direction in the wall-normal direction and the
$z$-direction is pointing upward.
The Leray-$\alpha$ simulations will be compared to the DNS database of \cite{Versteegh1999}.
The size of the domain is $L_x \times L_y \times L_z = H \times 6 H \times 12
H$. The unusually large domain size is required to prevent long-range correlations from
influencing the statistics \citep{Versteegh1998}.
The medium is air with a viscosity $\nu=1.0 \times 10^{-5}\ \text{m}^2/\text{s}$
, expansion coefficient $\beta = 3.3 \times 10^{-3}\ \text{K}^{-1}$ and a
Prandtl number $Pr=0.709$.
The distance $H$ between the plates is $H=0.2$ m and with a temperature
difference $\Delta \Theta = 2.7$ K with $T_0 = \Delta \Theta / 2$ and $T_1 = -
\Delta \Theta / 2$.
The Rayleigh number $\Ra = \beta g \Delta \Theta H^3 (\nu \kappa)^{-1}$ for this case is $\Ra=5 \times 10^6$.
The resolution for the coarse DNS is $64 \times 96 \times 192$, which as before
is chosen such that the grid is too coarse for accurate predictions but fine
enough to prevent that the results are dominated by numerical contamination.
The statistics have been collected over 25 typical turnover times.

In accordance with \cite{Versteegh1998, Versteegh1998a}, a body force is introduced which ensures a zero mass-flux. The advantage is that this suppresses slow transients, thereby reducing the required simulation time. In addition, for high $a$, the Leray-$\alpha$ model without body force the mass-flux becomes nonzero, which obfuscates comparison with other simulations.

The average velocity profile is shown in Fig.\ \ref{fig:lhc}b, normalized by
the free-fall velocity $\sqrt{\beta g \Delta \Theta H}$.
The coarse DNS overestimates the average flow velocity in the channel.
When increasing $a$, the velocity decreases and the flow profile is
approximately correctly predicted for $a=0.25$.
For $a=0.50$, the velocity is underpredicted.
At $a=1.00$, the velocity profile remains at roughly the same amplitude as
for $a=0.5$ but here a 'kink' can be observed in the mean velocity around
$x/H=0.2$, which is not present in the other simulations.
This kink is responsible for the shift of the shear production term 
$\av{\fl{\sm{u}} \fl{w}} \ddx{x} {\av{w}}$ from the center to the near-wall
region.

The results for $\Nu=\ddx{x}{\av{\Theta}} |_w H / \kappa \Delta \Theta$ as a
function of $a$ are shown in Fig.\ \ref{fig:lhc}c.
The increase in the average velocity does not influence $\Nu$ very much for
the coarse DNS.
The trend for increasing $a$ is that $\Nu$ decreases, similar to
Rayleigh-B\'{e}nard convection.
Around $a=0.25$, $\Nu$ is at its expected value, and above this value, $\Nu$ is
underpredicted.
The shear Reynolds number $\Re_\tau$ is defined as $\Re_\tau = u_\tau H / \nu$.
The influence of $a$ is relatively small for this flowcase (Fig.\
\ref{fig:lhc}d), with an underprediction of $\Re_\tau$ of 10\% at $a=1.00$.

The streamwise velocity variance profile $\av{\fl{w}\fl{w}}$ is shown in Fig.\ \ref{fig:lhc}e, nondimensionalized by $\beta g \Delta \Theta H$.
The coarse DNS overestimates the variance by 50\% in the bulk.
Here the effect of increasing $a$ is to decrease the variance for $a=0.25$.
As $a$ is increased further, the variance becomes practically constant in the
bulk ($a=0.50$), after which the maximum shifts from the center to the near-wall
region ($a=1.00$).

The equation of TKE for the side-heated vertical channel is given by
\begin{equation}
\label{eq:lhc_tke}
0 = 
\underbrace{\beta  g \av{\fl{w} \fl{\Theta}}
- \av{\fl{\sm{u}}\fl{w}} \ddx{x}{\av{w}}}_{\bP} - 
\underbrace{\nu
 \av{(\ddxj{\fl{u}_i})(\ddxj{\fl{u}_i}})}_{\bD} 
- \underbrace{\ddx{x}(\av{\fl{\sm{u}} e'} - \nu \ddx{x}{e} +
\av{\fl{u} \fl{p}})}_{\bT}. 
\end{equation}
For this flow case, there is production of TKE both by shear and 
by buoyancy.
Both components of $\bP$ are shown as a function of $z/H$ in
Fig. \ref{fig:lhc}f.
Here, it can be seen that the lack of resolution mainly affects the
shear-production term $-\av{\fl{\sm{u}} \fl{w}} \ddx{x}{\av{w}}$ through the
overestimation of the average velocity (Fig.\ \ref{fig:lhc}b).
The effect of increasing $a$ reduces the shear-production and the production is
estimated appropriately at $a=0.50$.
At $a=1.00$, the buoyancy production term changes little with respect to
$a=0.50$, but the shear production term changes dramatically.

\section{\label{par:discussion}Discussion}

The simulations show two major trends.
First, gradients at the wall (as reflected in integral quantities like $\Re_\tau$
and $\Nu$) tend to decrease.
Both for Rayleigh-B\'{e}nard convection and the side-heated vertical channel,
coarse DNS overpredicts wall gradients, and the Leray-$\alpha$ model can improve
results.
In the case that coarse DNS underpredicts gradients at the wall, as for the
plane channel flow, the Leray-$\alpha$ model does not improve the results.

Second, the variances tend to increase as a function of $\alpha$, specifically
at the low and intermediate wavenumbers.
In view of the model spectrum (Fig.\ \ref{fig:spectrum}), which predicts an increase in the variance at
wavenumbers larger than $k_\alpha$ only, this may be quite surprising.
However, this seems to be an intrinsic property of the Leray-$\alpha$ model. Simulations with the Leray-$\alpha$ model in absence of walls \citep{Geurts2006a, Graham2008} also results in enhanced variances and the low and intermediate wave lengths. The presence of walls does not change this property. However, the presence of a wall seems to enhance this feature of the model, due to an increase in the turbulent shear production term.  

In simulations with coarse grids, a significant part of the fluctuations are sub-grid, and separate modeling may be required \cite[e.g.][]{Carati2001}. Therefore, a pragmatic solution to remedy the increased variances may be to add some extra diffusion to the model. This may be done in several ways: 1) by using a simple eddy-viscosity model; 2) by using a dynamic Smagorinsky procedure; and 3) by using a spectral dissipation procedure.
It is clear that this procedure would have to be done with care, as one could easily overwhelm the Leray-$\alpha$ contribution to the subfilter stress. The tensor-diffusivity model (also called Clark model) \citep{Winckelmans2001}, is a good example of a model which significantly benefits from some additional dissipation.
This filter reproduces approximately 90\% of the subfilter stresses in \emph{a priori} studies . However, in \emph{a posteriori} studies, the tensor diffusivity model requires additional dissipation \cite{Winckelmans2001}.
Furthermore, extra dissipation may be unavoidable to be able to use
Leray-$\alpha$ with relatively coarse meshes.
In this study, the resolution was chosen such that coarse DNS simulations were
incapable of reproducing correct statistics but were not dominated by wiggles.
When even coarser grids would have been used, the absence of additional
dissipation leads to a deterioration of the results of the Leray-$\alpha$ model.


In this study we chose to keep the filter as close as possible to the original
formulation \citep{Cheskidov2005}.
The filtering requires inverting a Helmholtz equation for all three velocity components, which is computationally  expensive. In addition, the projection step to make $\sm{u}_i$ divergence-free requires solving another Poisson equation. Even though the computational overhead for solving a Poisson equation is minimized by using FFTs in the homogeneous directions, this procedure is more expensive than conventional SGS models.
To reduce the number of operations, one could use explicit filters
\citep{Geurts2006a, Liu2008} or apply
(\ref{eq:flt}) but with only one or two Jacobi iterations
\citep{Verstappen2007}.

An important question for future research is related to the influence of the filter type. It is not inconceivable that an explicit filter (i.e. a filter that does not require solving a Poisson equation, such as a top-hat filter) might perform better than the Helmholtz filter. Indeed, the Helmholtz filter suffers from $\flt(0)\neq0$ even for uniform $\alpha$, which violates the filtering framework \citep{vanReeuwijk2006}. A projection method was required to remedy this problem. In addition, the Helmholtz filter does not have compact support, causing the filtered velocity to be influenced relatively strongly by large fluctuations far away.

The inclusion of the buoyancy term in the Leray-$\alpha$ model has to be done with care.
In the direct simulations of Rayleigh-B\'{e}nard convection, the average
buoyancy production $\bP=\beta g \av{\fl{w}\fl{\Theta}}$ is constant in the
bulk, as $\bP$ depends directly on the turbulent heat flux
$\av{\fl{w}\fl{\Theta}}$ (which is constant in the bulk).
This is not the case for the Leray-$\alpha$ model, where a peak in $\bP$ is created near
the hydrodynamic boundary layer (Fig.\ \ref{fig:rb}e).
This change is a direct consequence of the modification of the turbulent heat
flux.
In the present formulation
(\ref{eq:leray2}-\ref{eq:theta2}), the
direct coupling between TKE production and heat-flux is broken, as the
equations for TKE and average heat-flux are given by (see also appendix
\ref{app:rbexact}):
\begin{gather}
\ddx{z}{\left(\av{\fl{\sm{w}} e'}
            + \av{\fl{p} \fl{w}}
            - \nu \ddx{z}{e} \right)}
            = \beta g \av{\fl{w} \fl{\Theta}}
            - \bD,  \\
\av{\fl{\sm{w}} \fl{\Theta}} - \kappa \ddx{z}{\av{\Theta}} = 
             \frac{\kappa \Delta \Theta}{H} Nu.
\end{gather}
The buoyancy production term is given by $\beta g \av{\fl{w}\fl{\Theta}}$, while the turbulent heat flux (which is constant in the bulk) is given by $\av{\fl{\sm{w}}\fl{\Theta}}$ (\ref{eq:rb_heatflux}).
The difference between $\av{\fl{w}\fl{\Theta}}$ and
$\av{\fl{\sm{w}}\fl{\Theta}}$ can be calculated by substituting $w = \sm{w} -
\ddxj{\alpha_j^2 \ddxj{\sm{w}}}$ (and therefore not correcting for
compressibility effects discussed in Sec.\ \ref{par:divergencefree})
into $\av{\fl{w}\fl{\Theta}}$ results in
\begin{equation}
\av{\fl{w}\fl{\Theta}} = \av{\fl{\sm{w}}\fl{\Theta}} -
                      \ddx{z}{(\alpha_z^2 \av{ \fl{\Theta}
\ddx{z}{\fl{\sm{w}}}})} +
                      \alpha_j^2 \av{ (\ddxj{\fl{\Theta}})(\ddxj{\fl{\sm{w}}})}.
\end{equation}
Interestingly, the two extra terms on the right-hand side correspond to terms
typically encountered in the transport equation of $\av{\fl{w}\fl{\Theta}}$
\citep{Hanjalic2002}.
The first is normally associated with the molecular diffusive transport, and the
second with the dissipative cross-correlation term, although the sign is
opposite here.

For buoyancy driven flows, the variation of $\alpha$ as a function of the wall-normal coordinate $z$ seems
be of importance for the variation of $\bP$ over the vertical in the bulk.
Indeed, the preliminary simulations with constant $\alpha$ (and free-slip
conditions for $\sm{u}$) did not have a variation of $\bP$ over the height
\citep{vanReeuwijk2005a}.
At $a=0.5$, the difference of $\bP$ between the peak value at the edge of the
boundary layer and the core is 10\% (Fig.\ \ref{fig:rb}e).
This is quite a large difference, and it seems worthwhile to explore whether this
effect can be circumvented.
The first and most obvious way is to experiment with different filters.
Another option is to modify the temperature equation.
Instead of holding on to the same advection operator for all transported
quantities, one could apply the modified advection operator $\sm{u}_j \ddxj{}$
to the momentum equations only.
For the other transported quantities, $u_j \ddxj{}$ could be used, with the
understanding that $u_j$ is also a regularized velocity.
This would directly restore the coupling between the turbulent heat-flux
and the buoyancy production.
However, the danger in a formulation like this may be an excess of variance at
small scales, which results in a high-wavenumber forcing contaminating the
simulation.

\section{\label{par:conclusions}Concluding remarks}

Numerical simulations of the Leray-$\alpha$ model have been carried out
for plane channel flow, Rayleigh-B\'{e}nard convection and the side-heated
vertical channel.
The simulations have been compared to DNS and coarse DNS.
In general, the simulations show two trends.
First, the viscous (and diffusive) wall region tends to thicken as a function of the filter width
parameter $a$, which causes wall gradients such as the shear stress
or the heat-flux to decrease.
When coarse DNS overpredicts wall gradients, such as for Rayleigh-B\'{e}nard
convection and (to a lesser extent) the side-heated vertical channel, the
Leray-$\alpha$ model can improve the results.
However, when the gradients are initially underestimated, such as for the plane
channel, results do not improve.
Here, additional wall-modeling may be needed to enhance turbulence levels in the
near-wall region.
Second, the variance at low and intermediate wavenumbers increases upon
increasing the filter size parameter $a$.
This leads to overpredicted variance in the velocity field which is undesired.

An important point for buoyancy driven flows is how to include a temperature
forcing into the Leray-alpha model.
Indeed, the intuitive extension
(\ref{eq:leray2}-\ref{eq:theta2})
causes the production of TKE by buoyancy to be no longer directly coupled
to the turbulent heat-flux.
For Rayleigh-B\'{e}nard convection, this leads to a varying TKE production in
the bulk (and as a function of $a$).

In this paper, the performance of the Leray-$\alpha$ model was assessed for three wall-bounded flows.
The Leray-$\alpha$ model was implemented in its original form, i.e. with the Helmholtz filter. 
This study indicates that, within this formulation, the potential of the Leray-alpha model is rather limited for wall-bounded flows. 
Indeed, the overpredicted variances, in particular the accumulation of energy on the low and intermediate wavenumbers pose a challenge for accurate predictions with the Leray-$\alpha$ model.
A study on alternative filter types, potentially remedying the aforementioned downsides of the Helmholtz filter, would be a valuable next step. 

\section*{Acknowledgments}
MvR thanks Prof.\ Darryl Holm for an inspiring stay at Imperial College in 2005.
We would like to thank Dr.\ Roel Verstappen and Prof.\ Bernard Geurts for
several useful discussions.
This work is part of the research programme of the Stichting voor Fundamenteel
Onderzoek der Materie (FOM), which is financially supported by the Nederlandse
Organisatie voor Wetenschappelijk Onderzoek (NWO).
The computations were sponsored by the Stichting Nationale Computerfaciliteiten
(NCF).

\appendix

\section{\label{app:rbexact}Changes in the dissipation rate for 
Rayleigh-B\'{e}nard convection}

The relations relating the volume-averaged dissipation rate of kinetic energy
$\hav{\bD}$ and temperature variance $\hav{\bD_{\Theta}}$ to the Rayleigh
number $\Ra$, the Prandtl number $\Pr$ and the Nusselt number $\Nu$  are given by
\citep{Grossmann2000, Siggia1994}
\begin{gather}
\label{eq:app_epsu}
\hav{\bD} = \frac{\nu^3}{H^4} \Ra (\Nu-1) \Pr^{-2}, \\
\label{eq:app_epsT}
\hav{\bD_{\Theta}} = \kappa \frac{\Delta \Theta^2}{H^2} \Nu,
\end{gather}
where $\hav{\bD}$ and and $\hav{\bD_\Theta}=\hav{\bD_{\Theta, \text{av}}}+
\hav{\bD_{\Theta, \text{fl}}}$ are defined as
\begin{gather*}
\hav{\bD}=\nu \hav{\av{(\ddxj{\fl{u}_i})(\ddxj{\fl{u}_i})}}, \\
\hav{\bD_{\Theta, \text{av}}}= 
       \kappa \hav{(\ddx{z}{\av{\Theta}})^2}, \\
\hav{\bD_{\Theta, \text{fl}}} = 
       \kappa \hav{\av{(\ddxj{\fl{\Theta}})(\ddxj{\fl{\Theta}})}}.
\end{gather*}
Here, $\hav{\cdot}$ represents the averaging over the entire fluid layer, and
the combination $\hav{\av{\cdot}}$ is equal to a volume- or ensemble-average,
when the system is ergodic.
Below we will derive the relations for $\hav{\bD}$ and $\hav{\bD_{\Theta}}$
for the Leray-$\alpha$ model.

The equations (\ref{eq:app_epsu}) and
(\ref{eq:app_epsT}) are obtained from the equations of
the vertical heatflux, turbulent kinetic energy and the two equations for
temperature variance.
In the case of the Leray-$\alpha$ model, these equations are given by
\begin{gather}
\av{\fl{\sm{w}} \fl{\Theta}} - \kappa \ddx{z}{\av{\Theta}} = 
             \frac{\kappa \Delta \Theta}{H} Nu, \\
\ddx{z}{\left(\av{\fl{\sm{w}} e'} + 
              \av{\fl{p} \fl{w}} - 
              \nu \ddx{z}{e} \right)}  = 
         \beta g \av{\fl{w} \fl{\Theta}} - \bD,  \\
\ddx{z}{\left( \av{\fl{\sm{w}} \fl{\Theta}}\ \av{\Theta} - 
               \kappa \ddx{z}{\frac{1}{2} \av{\Theta}\ \av{\Theta}}
        \right)}  = 
        \av{\fl{\sm{w}} \fl{\Theta}} \ddx{z}{\av{\Theta}} - 
        \bD_{\Theta, \text{av}}, \\
- \kappa \ddxsq{z}{\frac{1}{2} \av{\fl{\Theta} \fl{\Theta}}} = 
        - \av{\fl{\sm{w}} \fl{\Theta}} \ddx{z}{\av{\Theta}} - 
        \bD_{\Theta, \text{fl}},
\end{gather}
where, $e' = \fl{u}_i\fl{u}_i$ and $e = \av{\fl{u}_i\fl{u}_i}$.

Averaging these expressions over the height, we obtain
\begin{gather}
\label{eq:app_heatfluxav}
\hav{\av{\fl{\sm{w}} \fl{\Theta}}} = \frac{\kappa \Delta \Theta}{H} (\Nu-1) \\
\label{eq:app_epsufl}
\hav{\bD} =   \beta g \hav{\av{\fl{w} \fl{\Theta}}}    \\
\label{eq:app_epsTavg}
-\kappa \frac{\Delta \Theta^2}{H^2} Nu = 
           \hav{\av{\fl{\sm{w}} \fl{\Theta}} \ddx{z}{\av{\Theta}}} -
\hav{\bD_{\Theta, \text{av}}}\\
\label{eq:app_epsTfl}
0 =  - \hav{\av{\fl{\sm{w}} \fl{\Theta}} \ddx{z}{\av{\Theta}}}
     - \hav{\bD_{\Theta, \text{fl}}}
\end{gather}

Adding (\ref{eq:app_epsTavg}) and 
(\ref{eq:app_epsTfl}), the exact relation for
$\hav{\bD_\Theta}$, (\ref{eq:app_epsT}) is obtained.
However, this is not the case for the relation of the dissipation rate of TKE
(\ref{eq:app_epsu}).
Instead of the heat-flux based on the filtered velocity
$\av{\fl{\sm{w}}\fl{\Theta}}$, (\ref{eq:app_epsu}) contains
the heat-flux of the unfiltered velocity as $\av{\fl{w}\fl{\Theta}}$.
Substituting $w = \sm{w} - \ddxj{\alpha_j^2 \ddxj{\sm{w}}}$ (so not correcting
for compressibility effects), the two heat-fluxes are non-trivially coupled by
\begin{equation}
\label{eq:app_heatfluxunf}
\av{\fl{w}\fl{\Theta}} = \av{\fl{\sm{w}}\fl{\Theta}} -
                   \ddx{z}{\alpha_j^2 \av{ \fl{\Theta} \ddx{z}{\fl{\sm{w}}}}} +
                   \alpha_j^2 \av{ (\ddxj{\fl{\Theta}})(\ddxj{\fl{\sm{w}}})}.
\end{equation}
Interestingly, the two extra terms on the right-hand side correspond to terms
typically encountered in the transport equation for the turbulent heat-flux
\citep{Hanjalic2002}.
The first is normally associated with the molecular diffusive transport, and the
second with the dissipative cross-correlation term, although the sign is
opposite here.
Averaging (\ref{eq:app_heatfluxunf}) over the height and
substituting the result into (\ref{eq:app_epsufl}) using
(\ref{eq:app_heatfluxav}), gives that $\hav{\bD}$ is given
by
\begin{equation}
  \hav{\bD} = \frac{\nu^3}{H^4} \Ra (\Nu-1) \Pr^{-2} +
                \beta g \hav{\alpha^2 \av{
(\ddxj{\fl{\Theta}})(\ddxj{\fl{\sm{w}}})}}.
\end{equation}

\end{document}